\documentclass[prb,twocolumn,a4paper,floatfix,superscriptaddress,unsortedaddress,showpacs,showkeys]{revtex4-1}

\usepackage{dcolumn,amsmath,xspace}
\usepackage{graphicx}
\usepackage{epstopdf}

\newcommand{\degC}{\ensuremath{~^{\circ}\text{C }}}
\newcommand{\sixrt}{\ensuremath{(6\sqrt{3}\!\times\!6\sqrt{3})\text{R}30^\circ}~}

\begin{document}
\title{Symmetry breaking in commensurate graphene rotational stacking; a comparison of theory and experiment.}

\author{J. Hicks}
\author{M. Sprinkle}
\author{K. Shepperd}
\author{F. Wang}
\affiliation{The Georgia Institute of Technology, Atlanta, Georgia 30332-0430, USA}

\author{A. Tejeda}
\affiliation{Institut Jean Lamour, CNRS - Univ. de Nancy - UPV-Metz, 54506 Vandoeuvre les Nancy, France}
\affiliation{Synchrotron SOLEIL, L'Orme des Merisiers, Saint-Aubin, 91192 Gif sur Yvette, France}
\author{A. Taleb-Ibrahimi}
\affiliation{UR1 CNRS/Synchrotron SOLEIL, Saint-Aubin, 91192 Gif sur Yvette, France}

\author{F. Bertran}
\author{P. Le F\`{e}vre}
\affiliation{Synchrotron SOLEIL, L'Orme des Merisiers, Saint-Aubin, 91192 Gif sur Yvette, France}

\author{W.A. de Heer}
\affiliation{The Georgia Institute of Technology, Atlanta, Georgia 30332-0430, USA}

\author{C. Berger}
\affiliation{The Georgia Institute of Technology, Atlanta, Georgia 30332-0430, USA}
\affiliation{CNRS/Institut N\'{e}el, BP166, 38042 Grenoble, France}

\author{E.H. Conrad}
\affiliation{The Georgia Institute of Technology, Atlanta, Georgia 30332-0430, USA}

\begin{abstract}
Graphene stacked in a Bernal configuration ($60^\circ$ relative rotations between sheets) differs electronically from isolated graphene due to the broken symmetry introduced by interlayer bonds forming between only one of the two graphene unit cell atoms.  A variety of experiments have shown that non-Bernal rotations restore this broken symmetry; consequently, these stacking varieties have been the subject of intensive theoretical interest.  Most theories predict substantial changes in the band structure ranging from the development of a Van Hove singularity and an angle dependent  electron localization that causes the Fermi velocity to go to zero as the relative rotation angle between sheets goes to zero.  In this work we show by direct measurement that non-Bernal rotations preserve the graphene symmetry with only a small perturbation due to weak effective interlayer coupling.  We detect neither a Van Hove singularity nor any significant change in the Fermi velocity.  These results suggest significant problems in our current theoretical understanding of the origins of the band structure of this material.
 \end{abstract}
\vspace*{4ex}

\maketitle
\newpage

\section{Introduction\label{S:Intro}}
\begin{figure*}
\includegraphics[angle=0,width=16.0cm,clip]{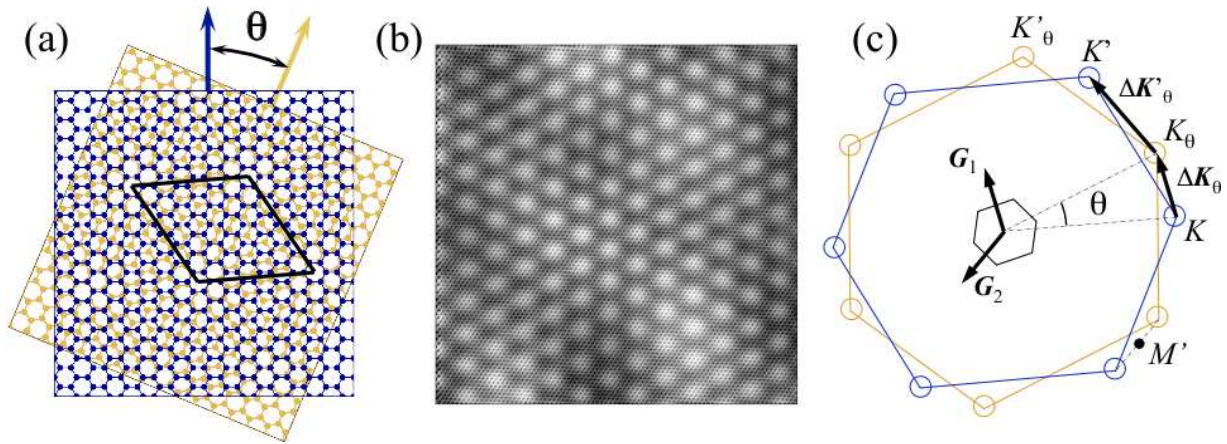}
\caption{(a) Schematic of the commensurate rotation of two graphene sheets.  The commensurate angle $\theta$ leads to a large supercell structure.  (b) a $200\!\times\!200\text{\AA}$ STM image of a $(m,n) = (4,5)$ supercell for C-face graphene sheets with a relative rotation of $\theta\!=\! 7.34^\circ$.\cite{Phil_Private} (c) The hexagonal Brillouin Zone of two graphene sheets with a relative rotation $\theta$.  The smaller Brillouin Zone of the commensurate superlattice (defined by reciprocal lattice vectors ${\bf G}_1$ and ${\bf G}_2$) is also shown.} \label{F:SuperCell}
\end{figure*}

Multilayer epitaxial graphene (MEG) grown on the SiC$(000{\bar1})$ (C-face) is now known to have a highly-ordered {\it non}-Bernal ({\it non}-{\it AB}) stacking, where adjacent graphene planes have commensurate relative rotations that are {\emph not}  $60^\circ$.\cite{Hass_PRL_08,Sprinkle_JphysD_10}  These commensurate rotations lead to large supercells [as shown in Fig.~\ref{F:SuperCell}(a)] that are seen in STM images of MEG films [see  Fig.~\ref{F:SuperCell}(b)].  These STM viewed supercells are often referred to as ``Moir\'{e}'' patterns. 

\begin{figure}
\includegraphics[width=7.9cm,clip]{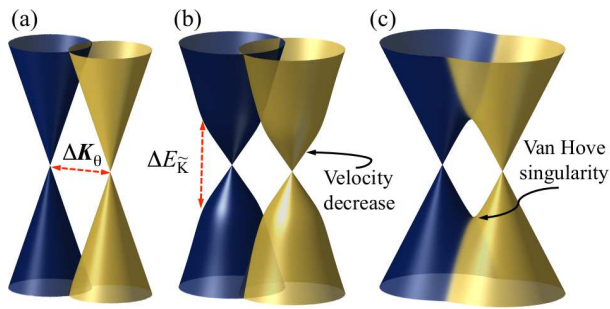}
\caption{(a) Schematic Dirac cones of two commensurately rotated graphene sheets, The cones are shifted by a vector $K_\theta$.  (b)  Velocity renormalization causes $v_F$ to be smaller than an isolated sheet making the slopes of the cones smaller in an energy window $\Delta E_{\tilde K}$ near the Dirac point. (c)  Predicted Van Hove singularity formed between the $K$- and $K_\theta$-points of the two rotated Dirac cones.} \label{F:Theory}
\end{figure}

Since the discovery of C-face MEG films, the effect of this new graphene stacking on the band structure of these films has been the subject of active experimental and theoretical study.  The earliest {\it ab-initio} calculations for a large-angle commensurate graphene bilayer rotation ($32.20^\circ$) predicted essentially no effect on the graphene band structure near the $K$-points of either graphene sheet.\cite{Latil_PRB_07,Hass_PRL_08}  In other words, the band structure consists of two independent but rotated Brillouin Zones (BZ) as shown in Fig.~\ref{F:SuperCell}(c) where nearby Dirac cones from the two sheets at $K$ and $K_\theta$ do not interact as shown in Fig.~\ref{F:Theory}(a). This prediction has been borne out in a number of experiments on MEG films. Electron transport,\cite{Berger06} infrared adsorption spectroscopy,\cite{Orlita_PRL_08} angle-resolved photoemmision (ARPES),\cite{Sprinkle_PRL_09,Sprinkle_PSSRRL_09} and scanning tunneling spectroscopy (STS)\cite{Miller_Science_09} all show that the graphene sheets in these films behave nearly identically to electronically-isolated graphene sheets.

Numerous theoretical treatments for rotated graphene bilayers, however, have predicted more dramatic effects in the band structure of these films.  Continuum approximations to a tight-binding model,\cite{LopesdoSantos_PRL_07}  as well as full tight-binding (TB) and {\it ab inito} calculations on small angle commensurate graphene bilayers,\cite{Mayou_NL_10}  predict a substantial decrease (renormalization) in the Fermi velocity, $v_F$, near the $K$-point (Dirac point).  In these calculations the interlayer interaction causes the wave functions in the graphene sheet to become highly localized for small relative rotations of the bilayer.\cite{Mayou_NL_10}  These localized states have a reduced $v_F$ that is exceptionally small when the relative rotation is less than $3^\circ$ -- less than $25\%$ of the value for an isolated graphene sheet.   This velocity reduction occurs in an energy window, $\Delta E_{\tilde{K}}$, defined as the energy difference between the Dirac point energy (i.e. at the $K$-point) and the energy where the two rotated Dirac cones cross (the $M'$-point) [see Figs.~\ref{F:SuperCell}(c) and \ref{F:Theory}(b)].\cite{Mayou_NL_10,Mele_TBP}   

In addition to the velocity renormalization, Lopes dos Santos, et al.\cite{LopesdoSantos_PRL_07} have also predicted that the interlayer interaction leads to an angle dependent Van Hove singularity in rotated bilayers.  The singularity appears where the Dirac cones from the two rotated layers cross at the $M'$-point [see Figs.~\ref{F:SuperCell}(c) and \ref{F:Theory}(c)]. 

MEG films offer a perfect platform for definitive tests of these predictions.  While MEG films have average relative rotations of $\sim\! 30^\circ$, much smaller relative rotations occur with significant frequency.\cite{Hass_PRL_08,Sprinkle_JphysD_10,Miller_Science_09}  Furthermore, these films are flat\cite{Hass_PRB_07}, extremely well-ordered,\cite{Hass_JPhyCM_08,Ming_PACS} and can be grown as thick multilayers or as thin as single sheets.  

In this work, we use high resolution ARPES to directly measure the band structure of commensurate rotated graphene sheets. These experiments conclusively show, despite the theoretical predictions for large reductions in $v_F$ and the development of a Van Hove singularity, that commensurately-rotated graphene sheets show no significant deviations from the linear band structure of graphene.  Measurements on many samples and many relative rotations, from both thick and thin films, show that the band structure of MEG films remains nearly identical to that of an isolated graphene sheet.  %These conclusive results can be reconciled with previous theoretical predictions and their limited supporting experimental evidence by considering the limitations of standard theoretical models and previous non-ideal experimental systems.

\section{Experimental\label{S:Exp}}
The substrates used in these studies were n-doped $n\!=\!2\times\!10^{18}\text{cm}^{-2}$ 6H-SiC. Samples were grown in a closed RF induction furnace using the Controlled Silicon Sublimation (CSS) method.\cite{Ming_PACS}  The samples were transported in air before introduction into the UHV analysis chamber.  Prior to ARPES measurements the graphene films were thermally annealed at 800\!\degC in UHV.  Sample thickness was measured by ellipsometry.\cite{Sprinkle_Thesis}  ARPES measurements were made at the Cassiop\'{e}e beamline at the SOLEIL synchrotron in Gif/Yvette. The high resolution Cassiop\'{e}e beamline is equipped with a modified Peterson PGM monochromator with a resolution $E/\Delta E \simeq 70000$ at 100 eV and 25000 for lower energies. The detector is a $\pm 15^{\circ}$ acceptance Scienta R4000 detector with resolution $\Delta E\!<\!1$meV and $\Delta k\!\sim\!0.01\text{\AA}^{-1}$ at $\hbar \omega\!=\!36$ eV. All measurements were carried out at 4K. The total measured instrument resolution is ($\Delta E\!<\!12$meV). 

\section{Results\label{S:Results}}
To understand subsequent ARPES data, we first review how commensurate rotations determine the band structure of rotated graphene.  Using the notation of Mele,\cite{Mele_PRB_10}  a commensurate rotation is determined by the supercell vector ${\bf a} = m{\bf t}_1+n{\bf t}_2$, where $m$ and $n$ are integers and ${\bf t}_1$ and ${\bf t}_2$ are the unit cell vectors of graphene.  The commensurate relative rotation of the two sheets is determined by integers $m$ and $n$;  $\cos{\theta} = (4mn+n^2+m^2)/2(m^2+n^2+mn)$.  In reciprocal space the commensurate supercell forms a small Brillouin Zone (BZ) defined by the two reciprocal lattice vectors ${\bf G}_1$ and ${\bf G}_2$  defined by;
\begin{subequations}
\begin{equation}
{\bf G}_1= \frac{m+n}{c}{\bf Q}_{g1} +  \frac{n}{c}{\bf Q}_{g2}\label{E:Q1}
\end{equation}
\begin{equation}
{\bf G}_2= -\frac{n}{c}{\bf Q}_{g1} +  \frac{m}{c}{\bf Q}_{g2},\label{E:Q2}
\end{equation}
\label{E:G-Vec}\end{subequations}
where $c\!=\!m^2\!+\!n^2\!+\!mn$ and ${\bf Q}_{g1}$ and ${\bf Q}_{g2}$ are the reciprocal lattice vectors of the unrotated graphene.  This leads to a supercell rotated by an angle $\phi$ relative to ${\bf Q}_{g1}$;  $\cos{\phi} = (m+n/2)/\sqrt{m^2+n^2+mn}$.  The reciprocal space picture then consists of two graphene BZs with a relative rotation $\theta$ as shown in Fig.~\ref{F:SuperCell}(c).  

Because the two sheets are commensurate, the $K$-points of the two BZs can be connected by linear combinations of commensurate supercell reciprocal lattice vectors, i.e. ${\bf \mathcal{G}}\! \equiv p{\bf G}_1\!+\!q{\bf G}_2$, where $p$ and $q$ are integers.\cite{Mele_TBP} Depending on the symmetry of the supercell either ${\bf K}\!-\!{\bf K}_\theta\!\equiv\!\Delta{\bf K}_\theta\!=\!{\bf \mathcal{G}} $ (SE-even symmetry\cite{Mele_PRB_10}) or ${\bf K}'\!-\!{\bf K}_\theta\!\equiv\!\Delta{\bf K}_\theta '\!=\!{\bf \mathcal{G}} $ (SE-odd symmetry\cite{Mele_PRB_10}), but not both simultaneously [see Fig.~ \ref{F:SuperCell}(c)].\cite{Mele_TBP}  The ${\bf \mathcal{G}}$ vectors are related to the rotation angle by

\begin{subequations}
\begin{equation}
|{\bf \mathcal{G}}|\!=\!|\Delta{\bf K}_\theta|\!=\!2K_{\Gamma K}\sin{\theta/2},\label{E:Gtheta(a)}
\end{equation}
\begin{equation}
|{\bf \mathcal{G}}|\!=\!|\Delta{\bf K}'_\theta|\!=\!2K_{\Gamma K}\sin{(60-\theta)/2}.\label{E:Gtheta(b)}
\end{equation}
\label{E:Gtheta}
\end{subequations}
Where $K_{\Gamma K}\!=\!|Q_g|/\sqrt{3}$.  Without considering interactions between sheets in a bilayer, the adjacent Dirac cones at $K$ and $K_\theta$ are simply interleaved cones as shown in Fig.~\ref{F:Theory}(a).  

Going beyond independent graphene sheets is the subject of many theoretical works.  For small rotation angles Lopes dos Santos, et al.\cite{LopesdoSantos_PRL_07} have made an analytic solution to the tight-binding model for small relative rotations.  They use a continuum approximation to describe the large range of interlayer bonding geometries in the large supercells that result from a relative rotation.  By assuming that the interlayer hopping has a long wavelength nature so that the coupling between different valleys ($K$ and $K'$) can be ignored, they treat the interlayer coupling as being uniform.  This calculation predicts two observables.  First,  the two nearby cones separated by $\mathcal{G}$ re-hybridize to form a distorted single cone with a Van Hove singularity (saddle point) halfway between the two $K$ and $K_\theta$-points  as shown in Fig.~\ref{F:Theory}(c). The second prediction is that the Fermi velocity slows to a renormalized value, $\tilde{v}_F$,  that scales with rotation angle approximately as,\cite{LopesdoSantos_PRL_07} 
\begin{equation}
\tilde{v}_F/ v_{Fo}\approx 1-9[\tilde{t}_\perp/\hbar v_{Fo}|\Delta {\bf K}_\theta|]^2  \label{E:VF}
\end{equation}
where $\tilde{t}_\perp$ is the interlayer hopping parameter for the rotated bilayer.  For small rotations angles $\tilde{t}_\perp$ is assumed to be independent of angle ($\tilde{t}_\perp\!\sim\!0.4t_\perp$, $t_\perp\!\sim\!0.3$eV).\cite{Neto_RevModPh_09}  While Eq.~\ref{E:VF} overestimates the reduction in $v_F$ below $\sim\! 2^\circ$, both full TB and {\it ab initio} calculations confirm the continuum approximation prediction that the renormalized velocity  goes to zero very fast at rotation angles less than $4^\circ$.\cite{Mayou_NL_10}  

Both the existence of a VHS and a renormalized $v_F$ can be tested by directly measuring the band structure of MEG films.  To make such comparisons, the band structure from rotated commensurate graphene sheets with known relative rotations must first be identified.  We now describe exactly how commensurate rotation angles are measured in ARPES.

 \begin{figure*}[hbtp]
\includegraphics[width=16.5cm,clip]{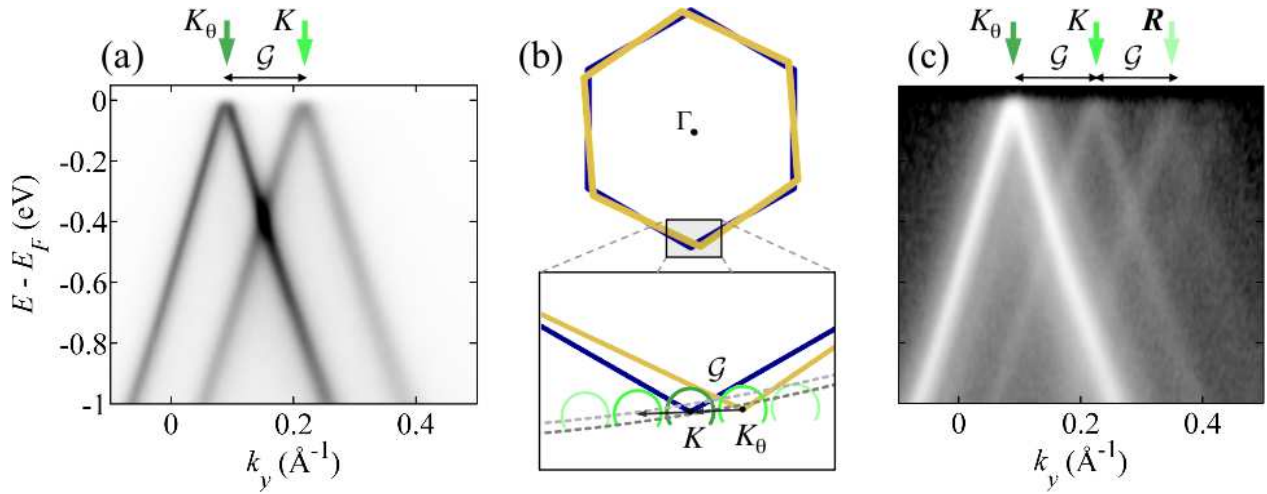}
\caption{(a) The band structure near the $K$- and $K_\theta$-points ($k_x=k_{\Gamma K}$) of two closely rotated planes (marked by arrows). $k_y$ is perpendicular to the $\Gamma\!-\!K$ direction.  The relative rotation angle between the two cones is  $4.2^\circ$.  (b) a schematic of the BZ of two rotated graphene sheets with a blowup near the $K$ and $K_\theta$ points. Two different detector planes (dashed arcs) are shown cutting through the Dirac cones and their replicas at $\pm n{\bf \mathcal{G}}$.  Note that the constant energy cuts through the cones are not full circles as described in text. (c) a similar ARPES scan of (a) but taken at a detector cut slightly closer to the $\Gamma$ point.  Data is plotted on a log scale to show the supercell $\mathcal{G}$-replicas band (${\bf R}$) (marked by a third arrow).  } \label{F:ReplicaCones}
\end{figure*}

Within the detector area of an ARPES measurement at $\hbar\omega\!=\!36$eV, Dirac cones from graphene sheets with relative rotations angles of $\pm 15^\circ$ are visible.  Some of the visible cones come from different uncorrelated rotated domains, i.e from a continuous graphene film that has an induced rotational boundary, $\Delta\phi$, caused by pleats in the graphene or by substrate step edges, etc. that are within the $30\mu\text{m}$ beam diameter.  
%For example a pleat in the top layer graphene sheet can produce two separate graphene regions to have a small relative rotation, $\phi$.  This will lead to two Dirac cones in the detector window that are separated by an angle $\Delta\phi$ but the two cones have no corresponding supercell.  On the other hand, two graphene sheets stacked on top of each other with a relative rotation will produce a pair of cones in ARPES where Eqs.~\ref{E:G-Vec} and \ref{E:Gtheta} are applicable so that the measured rotation angle is directly related to a ${\bf \mathcal{G}}$-vector of the supercell formed by the adjacent pair.
To distinguish the commensurate pairs formed by stacked rotated graphene sheets from incoherent pairs formed from pleats and steps, etc., we take advantage of the fact that in ARPES the photo-emitted electron can diffract from the local surface structure.  If there is no superlattice, as in the case of cones from two incoherent rotational domains, there will be {\bf no} diffraction.  On the other hand if the cones are from a coherent bilayer pair that form a superlattice, additional replica cones caused by diffraction will be visible in ARPES.  This diffraction effect has been nicely demonstrated for graphene grown on the Si-face of SiC where a \sixrt superstructure causes replica Dirac cones at ${\bf \mathcal{G}}$-vectors relative to the main graphene Dirac cones [see Refs.~\onlinecite{Bostwick_NPHYS_07} and \onlinecite{Bostwick_NJP_07}].  In the case of C-face graphene, the APRES signature of two stacked commensurately rotated graphene sheets will not only be the two primary Dirac cones but will also include replica (diffraction) cones positioned at $\pm {\bf \mathcal{G}}$ relative to either of the primary cones [ see Fig.~\ref{F:ReplicaCones}(b)].  

An example of how a commensurate pair is identified is shown in Fig.~\ref{F:ReplicaCones}.  Figure \ref{F:ReplicaCones}(a) shows a MEG band structure measured by ARPES near the graphene $K$- and $K_\theta$-point of two closely spaced Dirac cones ($\Delta k_y\!=\! 0.124\pm .005\text{\AA} $, ($k_x\!=\!k_{\Gamma K}$ ).  The two cones are identified as being part of a commensurate rotated pair as demonstrated in the log intensity plot of Fig.~\ref{F:ReplicaCones}(c).  The data is plotted in a log scale because the diffraction cross-section of the photo-electron is weak.  This makes the replica intensity significantly smaller compared to the primary cones. \cite{Bostwick_NJP_07}  The $k_x$ distance to the $\Gamma$ point in Fig.~\ref{F:ReplicaCones}(c) is slightly reduced relative to the image in Fig.~\ref{F:ReplicaCones}(a) to enhance the replica's intensity [see Fig.~\ref{F:ReplicaCones}(b) ].  The replica cone $R$ (marked by a third arrow in Fig.~\ref{F:ReplicaCones}(c)) has the same separation from the $K$ point as the $K$ point is from the $K_\Theta$ point.  Note that because the detector plane is not exactly perpendicular to the $\Gamma K$ direction and because constant energy cuts through the cones are only partial circles,\cite{Bostwick_NJP_07} replicas appear more intense in the $+k_y$ direction. Using Eq.~\ref{E:Gtheta}, the $\mathcal{G}$-vector connecting the pair of Dirac cones in  Fig.~\ref{F:ReplicaCones}(a) allows us to find the relative rotation of the two sheets to be $4.15\pm 0.07^\circ$.    Note that the $k_y$-width of the bands is resolution limited ($\sim 0.01\text{\AA}^{-1}$), indicating that the supercell domain size is well ordered over a length scale of 800\AA.  

We can be more restrictive in how we identify commensurate rotations.  In a commensurate pair, one of the graphene pairs must lie below the other.  At the photon energy used in these experiments, the electron mean free path, $\Lambda$, is only $3.4\text{\AA}$ (approximately the graphene interlayer thickness).\cite{Seah_SurfIA_79}  This means that one cone in a pair should have relative intensity that is $\sim\!40\%$ of the most intense cone in the pair.  For the analysis used in the work, we analyze only commensurate rotated pairs identified by both the existence of a related replica cone and that the two cones have the proper relative intensity for bilayers. 

\begin{figure}
\includegraphics[width=8.0cm,clip]{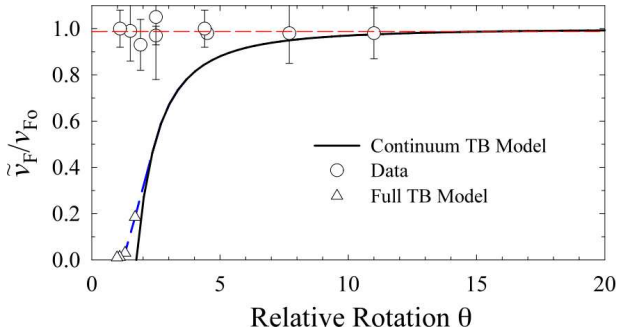}
\caption{Comparison of the measured $v_F$  ($\circ$) versus the predicted renormalized velocity  $\tilde{v}_F$ as a function of relative rotation angle.  Solid line is the approximate  $\tilde{v}_F$  from Ref.~[\onlinecite{LopesdoSantos_PRL_07}].  $\triangle$'s are the low angle corrected $\tilde{v}_F$  from a full TB  calculation.\cite{Mayou_NL_10}  Dashed line is the average velocity $\langle\!v_F\!\rangle$ for all measured Dirac cones.} \label{F:Exp_Vel}
\end{figure}

We have used ARPES spectra from similar cones pairs as that shown in Fig.~\ref{F:ReplicaCones} to investigate predictions for velocity renormalization $\tilde{v}_F$.  The Fermi velocity is derived from the slope of Dirac cones for relative rotations as low as $1.1^\circ$, for multiple samples, and for film thicknesses ranging from 3-10 layers.  All velocities were measured in an energy window between $E_D$ and the energy where the pair of cones cross (the $M'$-point in Fig.~\ref{F:SuperCell}]. The results are plotted in Fig.~\ref{F:Exp_Vel}. 

It is clear from Fig.~\ref{F:Exp_Vel} that no significant velocity changes are observed even for the smallest measured rotation angles, where the velocity has been predicted to be 50 times smaller than the  velocity for isolated graphene.  It must be pointed out that \textbf{no} cones, regardless of wether or not replica cones were visible, had any detectable velocity renormalization.  The average $\langle\!v_F\!\rangle$ for all measured cones from several samples and graphene thicknesses is $0.99\pm\! 0.05\!\times\!10^6$m/sec.  Within error bars,  the measured $\langle\!v_F\!\rangle$ is consistent with values obtained for exfoliated graphene\cite{Jiang_PRL_07} and from both IR measurements\cite{Orlita_PRL_08} and STS\cite{Miller_Science_09} on MEG films.  

\begin{figure}
\includegraphics[width=7.5cm,clip]{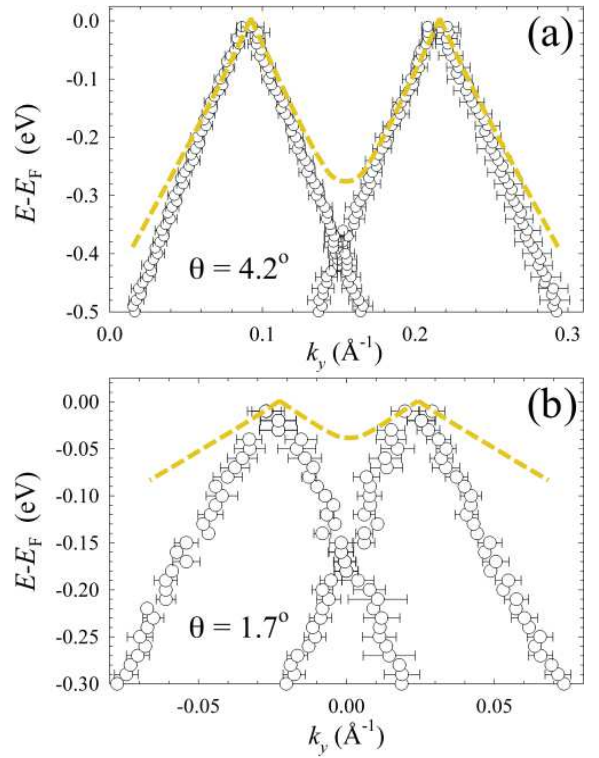}
\caption{A close up view of the band structure from (a) $4.2^\circ$ and (b) $1.7^\circ$ commensurate graphene pairs.  The dashed line is the predicted band structure with a Van Hove singularity from a continuum TB model.\cite{LopesdoSantos_PRL_07}  Note that the cones are hole doped by $\sim\!15$meV.} \label{F:WideCones}
\end{figure}

While the predicted velocity renormalization is a problem by itself, we are also able to show that no Van Hove singularity forms at the $M'$-point between the two cones [shown schematically in Fig.~\ref{F:Theory}(c)].  Figure \ref{F:WideCones} shows two detailed view near the crossing of two commensurately-rotated cones. The cones cross as straight lines with no singularity at the crossing (i.e, the $M'$-point).  The expected TB dispersion with VHS is drawn in Fig.~\ref{F:WideCones} for clarity. To within the experimental energy and momentum resolution, we can say that the crossing is an undistorted intersection of two linear bands.  Numerous samples and cones have been measured and no VHS are observed even for small angles like those in Fig.~\ref{F:WideCones}(b); precisely where the continuum model should be most appropriate.  Furthermore, no increase is observed in the integrated spectral density (proportional to the density of states) at the crossing that would indicate a change in $\vec{\bigtriangledown}_k(E)$ [this is discussed in more detail in Sec.~\ref{S:Discus}]. The experimental uncertainty places an upper limit on the size of a possible gap at the crossing to be less than 60meV.  While this is well below predicted values for a the Van Hove singularity of Lopes dos Santos, et al.,\cite{LopesdoSantos_PRL_07} the uncertainty does not rule out the small gaps predicted to occur for a class of rotations with weak interlayer coupling.\cite{Mele_TBP} 

%While many cones in our data set show these replicas, many other cones do not.  This is because the replica cones from large relative rotations  ($ > \pm 15^\circ$) will lie outside the detector window. This is consistent with X-ray measurements that show that the average relative rotation is $\sim 30^\circ$.\cite{Sprinkle_JphysD_10}  

%Either no replica cones will be visible or the replica cones will be separated in $k_y$ by a value not equal to the separation between cones $\Delta\tilde{k}$.\cite{note} 
 
\section{Discusion \label{S:Discus}}
The ARPES measurements presented in Sec.~\ref{S:Results} clearly demonstrate serious inconsistencies between the calculated and experimental band structure from commensurately rotated graphene sheets.   Both the predicted Van Hove singularity and renormalized Fermi velocity are not observed, even for small relative rotations.  While we cannot offer definitive reasons for the differences between current theory and these experiments, we can provide additional data that places limits on any model that may be used used to reconcile these differences.

First, it must be emphasized that surface effects cannot be the source of the inconsistency between theoretical predictions for $v_F$  and the values measured in these experiments.  This is because, even in thick MEG films ($10\!-\!40$ layers), infrared absorption experiments measure the same velocity.\cite{Orlita_PRL_08} 

A more important consideration is the valid energy range where renormalization is significant. Models differ in the energy range where the renormalized velocity prediction is expected to be valid.  More rigorous calculations, beyond first order expansions to a TB model, predict that at $k$-values larger than the first supercell BZ, the velocity should return to the isolated graphene value -- assuming no Van Hove singularity forms at the $M'$-point.\cite{Mele_TBP}  To ensure that the experimentally measured velocities are analyzed in an energy region valid for all models, we need to choose a minimum energy below the Dirac point where measured velocities are insured to be valid.  The predicted renormalized $\tilde{v}_F$ should be exact along the $\Gamma K$ direction within the first BZ of the commensurate supercell.\cite{Mayou_NL_10,Mele_TBP}  The energy window below the Dirac point is then set by  $\Delta{\bf K}_\theta$ and $\tilde{v}_F$; $\Delta E_{\Gamma K}\!=\!\hbar \tilde{v}_F| \Delta{\bf K}_\theta|$.  A more restrictive window would be to use the energy of the band at the cone crossing $M'$-point [see Fig.~\ref{F:SuperCell}(c) ], $\Delta E_{KM}$.  Using Eq.~\ref{E:VF}, the energy where the renormalized band intersects the $M'$-point would be:

\begin{equation}
\Delta E_{KM} \approx \frac{\hbar v_{Fo} |\Delta{\bf K}_\theta|}{2\sqrt{3}}[ 1-9(\tilde{t}_\perp/\hbar v_{Fo}|\Delta{\bf K}_\theta|)^2].  \label{E:E_BZ}
\end{equation} 

\begin{figure}
\includegraphics[width=7.5cm,clip]{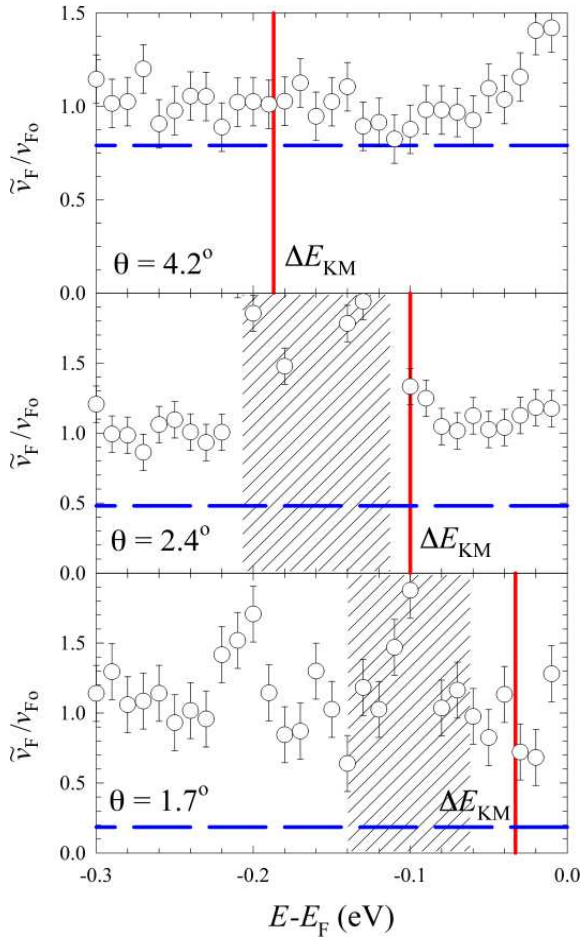}
\caption{Measured velocity, derived from the slope of the Dirac cones, as a function of energy.  The relative rotation angles are (a) $\theta\!=\!4.1^\circ$ and  (b) $\theta\!=\!2.4^\circ$.  The velocity is normalized by $v_{Fo}\!=\!10^6$m/sec.  The horizontal dashed line marks the TB prediction from Eq.~\ref{E:VF}.  Vertical line marks the predicted renormalized band energy $\Delta E_{KM}$ at the supercell $M$-point. The shaded region is the energy regime where the two cones cross and where $E(k)$ is subsequently difficult to accurately determine.} \label{F:VelvsE}
\end{figure}

We plot the normalized slope of the Dirac cones, $(1/\hbar)(\partial E/\partial k)(1/ v_{Fo})= \tilde{v}_F/v_{Fo}$, in Fig.~\ref{F:VelvsE} as a function of energy for rotated graphene pairs for three different rotation angles.  The energy $\Delta E_{KM}$ and the theoretical velocity reduction are marked for each rotation angle in Fig.~\ref{F:VelvsE}.  For the three angles shown, the measured velocity between $\Delta E_{KM}$ and $E_F$ is always higher than the predicted TB value and nearly equal to $10^6$m/sec.  We note that the change in slope close to the Dirac point ($\sim\!10$meV) should not be over interpreted.  The energy resolution  coupled with the $k$-resolution ($0.01\text{\AA}^{-1}$) and the distortion due to the Fermi-Dirac function near $E_F$ mean that it is difficult to accurately fit the position of the two converging $\pi$ bands.  A more important observation is that the measured velocity does not go to zero at $\Delta E_{KM}$ as it should if there was a Van Hove singularity at the $M'$-point.  These results show that if any significant velocity renormalization occurs, it is on an energy scale much smaller than current predictions. 

In contrast to this work, a singularity in the $dI/dV$ curves of STS experiments from CVD grown graphene deposited on graphite has been interpreted as evidence of a Van Hove singularity caused by rotated graphene sheets, although no corresponding velocity change has been reported.\cite{Li_NatPhys_10}  It is worth noting that there is a significant difference between MEG and graphene on graphite that makes the applicability of all the theoretical models discussed above tenuous for the graphene on graphite system.  First, the Moir\'{e} patterns of graphene on graphite are not necessarily due to simply sheet rotations. For example, surface graphite Moir\'{e} patterns are the result of a mechanically distorted top layer formed during cleaving of graphite.\cite{Pong_JPhysD_05}  Similar distortions and buckling are seen in depositing exfoliated graphene onto a support substrate. The consensus understanding is that these Moir\'{e} patterns are not caused by the relative rotation of a graphene sheets.\cite{Pong_JPhysD_05}  In other words, it is possible to have a Moir\'{e} pattern without rotations.  Furthermore, the Moir\'{e} corrugation amplitude in graphene on graphite films is typically $1\!-\!2\text{\AA}$  [see Pong and Durkan~\cite{Pong_JPhysD_05} and references therein].  This should be compared to MEG films where the corrugation is less than 0.2\AA,\cite{Hass_PRB_07,Miller_Science_09}  Height modulations that are more than a third of the graphene interlayer are certainly not part of the any of the models discussed above and therefore make a comparison between theories based on flat rotated sheets questionable.  

It has also been argued that the lack of a Van Hove singularity in MEG films implies that there is no interlayer coupling ($t_\perp\!\sim\!0$).\cite{Li_NatPhys_10}  This conjecture oversimplifies the problem for a number of reasons.  First, the coupling must be non-zero otherwise the films would simply delaminate in solution, which they do not.  Second, it is theoretically possible in TB to have a coupling and no Van Hove singularity.  Mele has recently shown that a certain class of rotation angles (those with SE-even rotations\cite{Mele_PRB_10} ), while having a non-zero interaction, do not have a singularity at the supercell $M'$-point.\cite{Mele_TBP} Third, recent STS experiments on the fine structure splitting of the zero-Landau level in MEG films indicate that third layer interaction are consistent with TB interlayer coupling parameters.\cite{Miller_Nphys_10} 

\section{Conclusion\label{S:Conclude}}
We have shown by directly measuring the band structure of multilayer epitaxial graphene that serious inconsistencies exist between the calculated and experimental band structure from commensurately rotated graphene sheets.  Both the predicted Van Hove singularity and the expected large reduction in the Fermi velocity, as the commensurate rotation angle between graphene sheets goes to zero, are not observed, at least not at the dramatic level predicted or in the energy window specified by current theories.  Given that these theoretical predictions are broad-based, encompassing both tight-binding and {\it ab intio} calculations, the inconsistency between theory and experiment suggests a fundamental process that is not included in current theoretical treatments.  It is possible that local strain fields caused by nearly overlapping $\pi$-bonds in the commensurate supercell lead to small relaxations that break some of the symmetry of the ideal rotated graphene pair.  While such speculations are interesting, they lie outside the scope of this work.  Nonetheless, the discrepancies pointed out this work suggest that a detailed look at these problems will be an important theoretical research avenue.

\begin{acknowledgments}
We wish to thank H. Tinkey and C. Johnson for their extensive help in analyzing ARPES data. This research was supported by the W.M. Keck Foundation, the Partner University Fund from the Embassy of France and the NSF under Grant No. DMR-0820382 and DMR-1005880.  We also wish to acknowledge the SOLEIL synchrotron radiation facilities and the Cassiop\'{e}e beamline. J. Hicks also wishes top acknowledge support from the NSF Graduate Research Fellowship Program.

\end{acknowledgments}

%%%%%%%%%%%%%%%%%%%%%%%%%%%%%%%%%%%%%%%%%%%%%%%%%%%%%%%%%%%%%%%%%%%%%%%%%%%%%%%
%%%%%%%%%%%%%%%%%%%%%%%%%%%%%%%%%%%%%%%%%%%%%%%%%%%%%%%%%%%%%%%%%%%%%%%%%%%%%%%

%\newpage

%Attenuation references

%\cite{Seah_SurfIA_79}
%$\Lambda = 3.4$\AA

%\cite{Cumpson_SurfIA_97}
%$\Lambda = 2.5$\AA

%\cite{Barrett_PRB_05}
%$\Lambda = 1.3$\AA

\end{document}